\documentclass[journal,comsoc]{IEEEtran}

\usepackage{amsmath}
\usepackage{amssymb}
\usepackage{balance}
\usepackage{cite}
\usepackage[T1]{fontenc}
\usepackage{graphicx}
\usepackage[utf8]{inputenc}
\usepackage{mathtools}
\usepackage{multirow}
\usepackage{url}
\usepackage{hyperref}
\begin{document}

\newcommand{\myauthor}{Kyeong Soo Kim}%
\newcommand{\mytitle}{Comments on ``On Clock Synchronization Algorithms for
  Wireless Sensor Networks Under Unknown Delay''}%

\title{\LARGE \mytitle}

\hypersetup{%
  pdfauthor={\myauthor},%
  pdftitle={\LARGE \mytitle},%
  pdfkeywords={},%
  pdfsubject={},%
}

\author{%
  Kyeong Soo Kim, \IEEEmembership{Member, IEEE}%
  \thanks{%
    K. S. Kim is with the Department of Electrical and Electronic Engineering,
    Xi'an Jiaotong-Liverpool University, Suzhou 215123, Jiangsu Province,
    P. R. China (e-mail: Kyeongsoo.Kim@xjtlu.edu.cn).%
  }%
}%


\maketitle

\begin{abstract}
  The generalization of the maximum-likelihood-like estimator for clock skew by
  Leng and Wu in the above paper is erroneous because the correlation of the
  noise components in the model is not taken into account in the derivation of
  the maximum likelihood estimator, its performance bound, and the optimal
  selection of the gap between two subtracting time stamps. This comment
  investigates the issue of noise correlation in the model and provides the
  range of the gap for which the maximum likelihood estimator and its
  performance bound are valid and corrects the optimal selection of the gap
  based on the provided range.
\end{abstract}


\section*{~}
As one of the three clock-synchronization algorithms studied for wireless sensor
networks (WSNs) under unknown delay \cite{leng10}, Leng and Wu proposed the
generalization of the maximum-likelihood-like estimator (MLLE) proposed by Noh
\textit{et al}. \cite{noh07:_novel}. To overcome the drawback of the MLLE that
it can utilize only the time stamps in the first and the last of $N$ message
exchanges, they extend the gap $\alpha$ between two subtracting time stamps from
$N{-}1$ to a range of $\left[1,\ldots,N{-}1\right]$ so that the generalized MLLE
can take into more time stamps in estimating clock skew.

Specifically, the time stamps in two-way message exchanges are modeled as
\cite[Eqs.~(1)~and~(2)]{leng10}
\begin{align}
  \label{eq:T1T2}
  T_{2,i} & = \beta_{1} T_{1,i} + \beta_{0} + \beta_{1} \left(d + X_{i}\right) \\
  \label{eq:T3T4}
  T_{3,i} & = \beta_{1} T_{4,i} + \beta_{0} - \beta_{1} \left(d + Y_{i}\right)
\end{align}
where $\beta_{0}$ and $\beta_{1}$ denote the clock offset and clock skew of the
child node $S$ with respect to the parent node $P$, respectively; $d$ represents
the fixed portion of one-way propagation delay, while $X_{i}$ and $Y_{i}$ are
its variable portions (see Fig.~1 of \cite{leng10}). Based on \eqref{eq:T1T2}
and \eqref{eq:T3T4}, they construct new sequences
$D_{r,j}{\triangleq}T_{r,\alpha+j}{-}T_{r,j}$ ($j{=}1,\ldots,N{-}\alpha$ and
$r{=}1,2,3,4$) and model them as follows \cite[Eqs.~(10)~and~(11)]{leng10}:
\begin{align}
  \label{eq:D1D2}
  D_{2,j} & = \beta_{1} D_{1,j} + \beta_{1} \left(X_{\alpha+j} - X_{j}\right) \\
  \label{eq:D3D4}
  D_{3,j} & = \beta_{1} D_{4,j} - \beta_{1} \left(Y_{\alpha+j} - Y_{j}\right)
\end{align}
for $j{=}1,\ldots,N{-}\alpha$. Noting that
$\left(X_{\alpha+j}{-}X_{j}\right){\sim}\mathcal{N}(0,2\sigma^{2})$ and
$\left(Y_{\alpha+j}{-}Y_{j}\right){\sim}\mathcal{N}(0,2\sigma^{2})$ because
$X_{j}$ and $Y_{j}$ are i.i.d. zero-mean Gaussian random variables with variance
$\sigma^{2}$, they obtain the maximum-likelihood estimator (MLE) for $\beta_{1}$
given by \cite[Eq.~(13)]{leng10}
\begin{equation}
  \label{eq:skew_est}
  \hat{\beta}_{1} = \dfrac{1}{\hat{\theta}_{1}}
  = \dfrac{\sum_{j=1}^{N-\alpha}\left(D_{2,j}^{2}+D_{3,j}^{2}\right)}{\sum_{j=1}^{N-\alpha}\left(D_{1,j}D_{2,j}+D_{4,j}D_{3,j}\right)} .
\end{equation}

The major problem in the derivation of the MLE for $\beta_{1}$ given in
\eqref{eq:skew_est} is that, even though $X_{j}$ and $Y_{j}$ are i.i.d. Gaussian
random variables, the noise components $\left(X_{\alpha+j}{-}X_{j}\right)$ and
$\left(Y_{\alpha+j}{-}Y_{j}\right)$ are not in general: For
$m,n{\in}\left\{1,\ldots,N-\alpha\right\}$ and $m{\neq}n$,
\begin{align}
  \label{eq:noise_correlation}
  \MoveEqLeft
  \operatorname{E}\left[\left(X_{\alpha+m}-X_{m}\right)\left(X_{\alpha+n}-X_{n}\right)\right]
  \notag \\
& = -\operatorname{E}\left[X_{\alpha+m}X_{n}\right] -
  \operatorname{E}\left[X_{m}X_{\alpha+n}\right] \notag \\
& = \begin{cases}
  -\sigma^{2}, & \mbox{if } \alpha=\left|m-n\right| \\
  0, & \mbox{otherwise}
\end{cases} ,
\end{align}
and the same goes for $\left(Y_{\alpha+j}{-}Y_{j}\right)$. Note that, if the
noise components are independent one another as claimed in \cite{leng10}, the
expectation in \eqref{eq:noise_correlation} must be zero.

The consequence of \eqref{eq:noise_correlation} is that $\alpha$ should be
greater than $\frac{N-1}{2}$, i.e.,
\begin{align}
  \label{eq:gap_range}
  \alpha \in \left\{\left\lfloor\frac{N}{2}\right\rfloor,\ldots,N-1\right\}
\end{align}
in order to maintain the validity of the derivation of the MLE for $\beta_{1}$
\cite[Eq.~(13)]{leng10} and its performance bound \cite[Eq.~(29)]{leng10}: If
$\alpha{\leq}\frac{N-1}{2}$, there exists at least one pair of $m$ and $n$
satisfying $\alpha=\left|m-n\right|$ so that the noise components are no longer
independent one another. For example, let $n$ be 1. Then $m{=}\alpha{+}1$
satisfies the said condition. Because $\alpha{\leq}\frac{N-1}{2}$ and
\[
m = \alpha + 1 \leq \dfrac{N-1}{2} + 1 = \dfrac{N+1}{2} = N - \dfrac{N-1}{2}
\leq N - \alpha ,
\]
$m$ belongs to
$\left\{1,\ldots,N-\alpha\right\}$.

\begin{figure}[!tb]
  \centering
  \includegraphics[width=\linewidth]{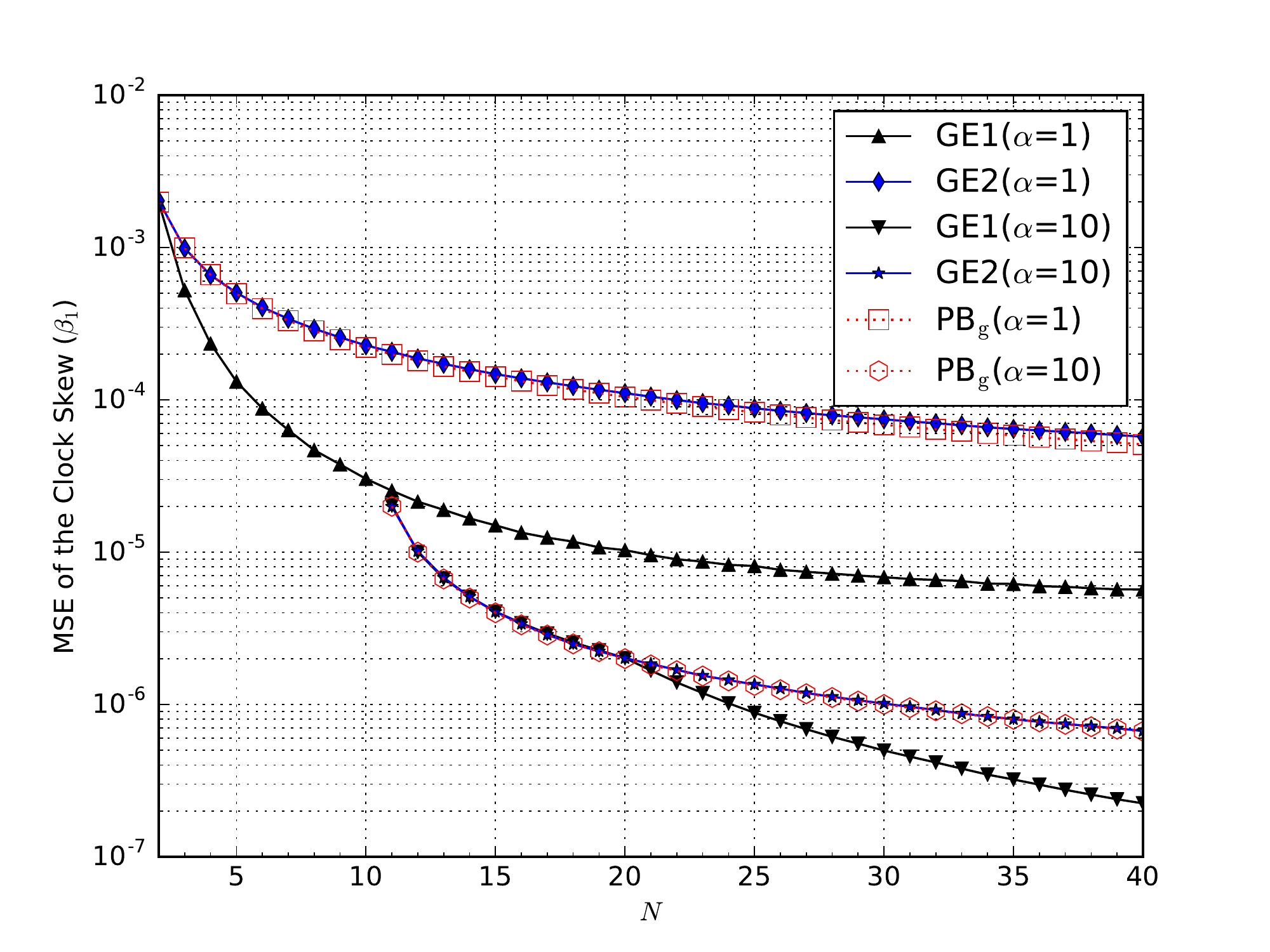}
  \caption{Effect of noise correlation on the MSE of estimated clock skew.}
  \label{fig:noise_correlation_effect}
\end{figure}

Fig.~\ref{fig:noise_correlation_effect} clearly shows the effect of the noise
correlation on the mean square error (MSE) of estimation of clock skew and the
relationship between $\alpha$ and $N$ when SNR{=}30 dB and $H{=}G{=}10$. In the
figure, GE1 denotes the simulation results of the generalized MLLE for time
stamps and resulting sequences generated according to the original models of
\eqref{eq:T1T2} through \eqref{eq:D3D4}; GE2, on the other hand, denotes the
results for the time sequences in \eqref{eq:D1D2} and \eqref{eq:D3D4} with the
noise components $\left(X_{\alpha+j}-X_{j}\right)$ and
$\left(Y_{\alpha+j}-Y_{j}\right)$ replaced by two newly-generated
i.i.d. zero-mean Gaussian random variables with variance
$2\sigma^{2}$.\footnote{It does not correspond to any model of two-way message
  exchanges and is given just for the purpose of comparison.}

If $\alpha$ is greater than $\frac{N-1}{2}$, we can see that the results of GE1
closely match with the performance bounds (i.e., PB$_{\rm g}$) because there is
no issue of noise correlation; for example, when $\alpha$ is 10, the results of
GE1 match with the performance bounds for $N$ up to 20. Compared to the results
for GE1, the results for GE2 of a fictitious model show that they can attain the
performance bounds irrespective of the value of $\alpha$ because there is no
issue of noise correlation at all. It is interesting, though, that the results
of GE1 for $\alpha\leq\frac{N-1}{2}$ show even better performance than the
performance bounds.

With the valid range of $\alpha$ given by \eqref{eq:gap_range}, the selection of
the optimal $\alpha$ given in Eqs.~(32)~and~(33) of \cite{leng10} should be
modified accordingly. Because $\Phi(\alpha_{r})$ in Eq.~(32) of \cite{leng10} is
concave downward for the whole range of real-valued
$\alpha_{r}{\in}\left[\left\lfloor\frac{N}{2}\right\rfloor,N{-}1\right]$,
$\alpha^{*}_{r}$ in Eq.~(33) of \cite{leng10} is now simplified as
follows\footnote{See \cite[Appendix~A]{leng10} for details.}:
\begin{align}
  \alpha^{*}_{r} = \dfrac{1}{3}N + \sqrt{\dfrac{1}{9}N^{2} - \dfrac{2\beta_{1}^{2}\sigma^{2}}{\beta_{1}^{2}H^{2}+G^{2}}}
\end{align}

\balance



\begin{thebibliography}{1}
\providecommand{\url}[1]{#1}
\csname url@samestyle\endcsname
\providecommand{\newblock}{\relax}
\providecommand{\bibinfo}[2]{#2}
\providecommand{\BIBentrySTDinterwordspacing}{\spaceskip=0pt\relax}
\providecommand{\BIBentryALTinterwordstretchfactor}{4}
\providecommand{\BIBentryALTinterwordspacing}{\spaceskip=\fontdimen2\font plus
\BIBentryALTinterwordstretchfactor\fontdimen3\font minus
  \fontdimen4\font\relax}
\providecommand{\BIBforeignlanguage}[2]{{%
\expandafter\ifx\csname l@#1\endcsname\relax
\typeout{** WARNING: IEEEtran.bst: No hyphenation pattern has been}%
\typeout{** loaded for the language `#1'. Using the pattern for}%
\typeout{** the default language instead.}%
\else
\language=\csname l@#1\endcsname
\fi
#2}}
\providecommand{\BIBdecl}{\relax}
\BIBdecl

\bibitem{leng10}
M.~Leng and Y.-C. Wu, ``On clock synchronization algorithms for wireless sensor
  networks under unknown delay,'' \emph{{IEEE} Trans. Veh. Technol.}, vol.~59,
  no.~1, pp. 182--190, Jan. 2010.

\bibitem{noh07:_novel}
K.-L. Noh, Q.~M. Chaudhari, E.~Serpedin, and B.~W. Suter, ``Novel clock phase
  offset and skew estimation using two-way timing message exchanges for
  wireless sensor networks,'' \emph{{IEEE} Trans. Commun.}, vol.~55, no.~4, pp.
  766--777, Apr. 2007.

\end{thebibliography}

\end{document}